\documentclass[twocolumn,showpacs,preprintnumbers]{revtex4}
\usepackage{amssymb}
\usepackage{amsmath}
\usepackage{graphicx}
\usepackage{dcolumn}
\usepackage{bm}

\begin{document}

\title{Shock propagation and stability in causal dissipative hydrodynamics}
\author{G.S.Denicol, T. Kodama, T. Koide, and Ph. Mota}
\address{Instituto de F\'{\i}sica, Universidade Federal do Rio de Janeiro, C. P.
68528, 21945-970, Rio de Janeiro, Brasil}

\begin{abstract}
We studied the shock propagation and its stability with the causal
dissipative hydrodynamics in 1+1 dimensional systems. We show that the
presence of the usual viscosity is not enough to stabilize the solution.
This problem is solved by introducing an additional viscosity which is
related to the coarse-graining scale of the theory.
\end{abstract}

\maketitle


\section{Introduction}

It is by now widely accepted that the basic features of collective motions
in relativistic heavy-ion collisions can be well described as those of a
hydrodynamical motion of an (almost) ideal fluid \cite{GeneralHydro}.
Several studies on the effects of viscosity are available in literatures and
support such a vision \cite{1st-a,1st-b,2nd,DKKM}. In addition, the
possibility of the existence of the lower bound of the shear viscosity
coefficient has been discussed by assuming the AdS/CFT correspondence \cite%
{LowViscosity}.

However, strictly speaking there are still several open questions in
hydrodynamic approaches of the heavy-ion collisions. 
Hydrodynamic observables are not so restrictive to determine uniquely many
unknown factors, such as initial condition, equation of state, dissipation
mechanisms, and so on. In particular, even the theory of a relativistic
dissipative fluid is not yet well understood. Therefore, in order to
conclude that the matter created in relativistic heavy-ion collisions really
behaves just as an ideal fluid, we need to investigate the effect of
dissipation more carefully.

The difficulty of the construction of the relativistic dissipative
hydrodynamics is because of the problem of acausality and instability: a
naive relativistic generalization of the Navier-Stokes equation present an
infinite propagation speed of pulse signals and the solution is unstable for
small perturbations. To solve this problem, one should, for example, take
into account memory effects by introducing a relaxation time \cite{DKKM}.

So far, there are several attempts to study the effect of dissipation to
relativistic heavy-ion collisions by implementing the 1+1 and 2+1
dimensional calculations \cite{2nd}. In these studies, they mainly deal with
cases where the deviation of the ideal fluid is rather small. On the other
hand, there is an evidence that the bulk viscosity becomes large in the
vicinity of the critical point\cite{Karsh}. In addition, to know the
limitation of the theory itself, we should investigate the behavior of
solutions for large viscosity as well. In particular, for higher energies
such as in the LHC regime, we expect that the relative importance of the
viscosity becomes significant. This point is essential since the problem of
acausality or instabilities is directly related with the size of viscosity
and the relaxation time mentioned above.

Other interesting aspects of viscous fluid dynamics appear when
discontinuities emerge during the time evolution or already exist in the
initial condition. In the usual application of hydrodynamics, only very
smooth initial distributions, both in energy and velocity, have been
applied. In such cases, the fluid profile usually remains smooth in time and
no special attention is required for the treatment of discontinuities.

However, it is sometimes necessary to discuss the extreme cases which
involve discontinuities. For example, the Landau type initial condition is
often discussed as an interesting possibility of meson production mechanism
with null initial velocity in the p-p collisions and A-A collisions \cite%
{LandauExp}. Furthermore, the shock phenomena, which are typical
discontinuous propagation of hydrodynamical variables, also may occur in the
heavy-ion collisions by high energy jet propagations in the QGP \cite{Mach,
1st-b}. We also expect a formation of shock wave in the region near the
coexisting phase, if the QCD-hadron phase transition is of first order,
where the velocity of sound vanishes (or becomes very small).\textbf{\ }
Shock phenomena in relativistic heavy ion collisions, if any, are
particularly interesting since they would furnish genuine hydrodynamical
signals.

It is well-known that to deal with dynamical discontinuities, such as
shocks, is not a simple problem. We have to introduce some specific
techniques such as the Godunov method or artificial viscosity
(pseudo-viscosity) \cite{Artificial} to achieve physically meaningful
results interpolating smoothly the discontinuity. For a causal dissipative
hydrodynamics, there are a few works where the dynamics of shock
discontinuities are discussed \cite{JP,Ruggeri}. However, to the author's
knowledge, detailed numerical study of a causal dissipative fluid dynamics
involving discontinuities has not yet been done.

In this paper, we study in detail the dynamics of a viscous fluid. For the
sake of simplicity, we concentrate ourselves to 1+1 dimensional systems. The
basic objective is to analyze the problem of instabilities associated with
causality through several numerical examples, in particular, which contain
discontinuities or shock phenomena. We also pay attention to the origin of
the so-called \textquotedblleft artificial viscosity\textquotedblright\ for
the numerical calculations of shock phenomena in the framework of causal
viscous hydrodynamics. We investigate its role and interpret its origin in
terms of the scale of the coarse graining introduced in the hydrodynamical
theory. To clarify this point, we use the smoothed particle hydrodynamics
(SPH) as the numerical method instead of the commonly used space-fixed grid
methods. Furthermore, having in mind the collective QGP motion in the LHC
energy regime, we restrict ourselves to systems described by an equation of
state of a baryon-number free, massless relativistic gas.

In the next section, we briefly review the causal dissipative hydrodynamics
and describe how the memory function is introduced. To check causality and
stability of our theory, we discuss the dispersion relation for the
propagation of a perturbative plane-wave. In Sec. III, we describe the SPH
method applied to our problem which introduces the coarse-graining scale $h$%
. In Sec. IV, we show several examples to reveal the effects of viscosity,
in particular, the case of steady shock wave propagation induced by a large
pressure gradient in the initial condition. In the first example, we show
the universal relation between pressure and viscosity in the fluid expansion
to vacuum. This relation is satisfied independently of initial conditions
and equation of states. In these examples, some quick oscillating modes
appear in the dynamics involving shocks, leading to instabilities of the
numerical solutions. That is, the normal viscosity does not necessarily damp
all the high frequency modes. The presence of such modes indicates the
necessity of a new ingredient for the theory to be physically consistent. In
Sec. V, we introduce an additional viscosity, having a different scale of
the memory function associated with the coarse graining size, which solves
this problem. We show several examples where the dynamics of shock phenomena
are described satisfactorily in this scheme. Finally in Sec.VI, we summarize
our work on the analysis of a causal dissipative hydrodynamics in 1+1 and
discuss the problems still open in such theories.

\section{Relativistic Dissipative Hydrodynamics}

Various theories have been proposed to incorporate dissipation consistent
with causality and stability; the divergence type theory \cite{muller}, the
Israel-Stewart theory \cite{II} and its extension based on the extended
irreversible thermodynamics \cite{Jou}, Carter's theory \cite{carter},
\"Ottinger-Grmela formulation \cite{OG} and the memory function method \cite%
{DKKM}.

Here we briefly review the memory function method \cite{DKKM} to obtain a
causal dissipative hydrodynamics for 1+1 dimensional case. That is, we
ignore the motion in the transverse direction and concentrate on the
longitudinal dynamics.

As was mentioned in the introduction, we just consider the case of vanishing
baryon chemical potential. In this case, the hydrodynamical equation of
motion can be written only as the conservation of the energy-momentum
tensor, 
\begin{equation}
\partial _{\mu }T^{\mu \nu }=0.  \label{divTmunu}
\end{equation}%
together with the thermodynamical relations. We use the Landau definition
for the local rest frame and assume, as usual, that the thermodynamic
relations are valid in this frame. Then the energy-momentum tensor is
expressed as 
\begin{equation}
T^{\mu \nu }=\left( \varepsilon +p+\Pi \right) u^{\mu }u^{\nu }-\left( p+\Pi
\right) g^{\mu \nu },  \label{NeTmunu}
\end{equation}%
where, $\varepsilon $, $p$, $u^{\mu }$ and $\Pi $ are, respectively, the
energy density, pressure, four velocity and bulk viscosity.

In the presence of these irreversible currents, the entropy is not conserved
anymore. Instead, from Eq.(\ref{divTmunu}), we have \cite{LL} 
\begin{equation}
\partial _{\mu }\sigma ^{\mu }=-\frac{1}{T}\Pi \partial _{\mu }u^{\mu },
\label{s-current}
\end{equation}%
where the entropy four-flux is identified by Landau and Lifshitz as 
\begin{equation}
\sigma ^{\mu }=su^{\mu }.  \label{sigma}
\end{equation}

In irreversible thermodynamics, it is interpreted that entropy production is
the sum of the products of thermodynamic forces and irreversible currents.
From Eq. (\ref{s-current}), we define the thermodynamic force as 
\begin{equation}
F=\partial _{\alpha }u^{\alpha }.
\end{equation}%
To satisfy the second law of thermodynamics locally, and hence the
positiveness of the entropy production, Landau proposed that the
irreversible current should be proportional to the thermodynamic force \cite%
{LL}, 
\begin{equation}
\Pi =-\zeta F=-\zeta \partial _{\alpha }u^{\alpha },  \label{viscous}
\end{equation}%
where $\zeta $ is the viscosity coefficient. However, it is known that the
derived equations have the problem of acausality and instability \cite%
{Hiscock,kouno,Gio}. To solve these difficulties, we introduce a memory
effect to the irreversible current by using a memory function\cite{DKKM}.
One of the simplest forms of the memory function is 
\begin{equation}
G\left( \tau ,\tau ^{\prime }\right) \rightarrow \frac{1}{\tau _{R}\left(
\tau ^{\prime }\right) }e^{-\int_{\tau ^{\prime }}^{\tau }\frac{1}{\tau
_{R}\left( \tau ^{\prime \prime }\right) }d\tau ^{\prime \prime }}.
\end{equation}%
The relaxation time $\tau _{R}(\tau )$ is, in general, a function of the
local proper time $\tau =\tau \left( \vec{r},t\right) $ through the
thermodynamical quantities. Then, the irreversible current is modified as
follows; 
\begin{equation}
\Pi \left( \tau \right) =-\int_{\tau _{0}}^{\tau }d\tau ^{\prime }G\left(
\tau ,\tau ^{\prime }\right) \zeta \partial _{\alpha }u^{\alpha }\left( \tau
^{\prime }\right) +e^{-(\tau -\tau _{0})/\tau _{R}}\Pi _{0},
\label{Integrals}
\end{equation}%
where $\Pi _{0}$ is the initial value given at $\tau _{0}$.

Because of the modification of the relation between the irreversible
currents and the thermodynamic forces, the algebraic positivity of the
second law of thermodynamics is not satisfied. However, we checked that the
second law of thermodynamics is not violated in all examples discussed in
this paper numerically. It is worth mentioning that the algebraic positivity
is not automatically satisfied even in the Israel-Stewart theory (See the
discussion below Eq. (2.31) in \cite{II}). To solve this problem, the
concepts of thermodynamics have to be extended. See \cite{Jou} for details.

The integral expression (\ref{Integrals}) are equivalent to the following
differential equation, 
\begin{equation}
\Pi =-\zeta \partial _{\alpha }u^{\alpha }-\tau _{R}\frac{d\Pi }{d\tau },
\end{equation}%
where $d/d\tau =u^{\mu }\partial _{\mu }$ is the total derivative with
respect to the proper time. In the 1+1 dimensional case discussed here, the
equations derived above are equivalent to those of the Israel-Stewart
theory. Thus, the conclusions in this paper is applicable also to the
Israel-Stewart theory.

\subsection{Propagation speed of signals}

We discuss the propagation speed of the 1+1 dimensional system. For this
purpose, we consider a small perturbation of the three independent variables
in the form of a plane wave, 
\begin{equation*}
\left( 
\begin{array}{c}
\delta \epsilon \\ 
\delta U^{1} \\ 
\delta \Pi%
\end{array}%
\right) \propto e^{i\omega t-ikx},
\end{equation*}%
propagating in a hydrostatic equilibrated background. Then, the linearized
hydrodynamic equation for these perturbations should satisfy 
\begin{equation}
\left( 
\begin{array}{ccc}
i\omega & -ik(\epsilon +p) & 0 \\ 
\alpha (-ik) & i\omega (\epsilon +p) & -ik \\ 
0 & -ik\zeta & 1+\tau _{R}\gamma i\omega%
\end{array}%
\right) \left( 
\begin{array}{c}
\delta \epsilon \\ 
\delta U^{1} \\ 
\delta \Pi%
\end{array}%
\right) =0,  \label{perturb}
\end{equation}%
where $\alpha \equiv dp/d\varepsilon $ is square of the velocity of sound in
the absence of the viscosity.

\begin{figure}[tbp]
\includegraphics[scale=1]{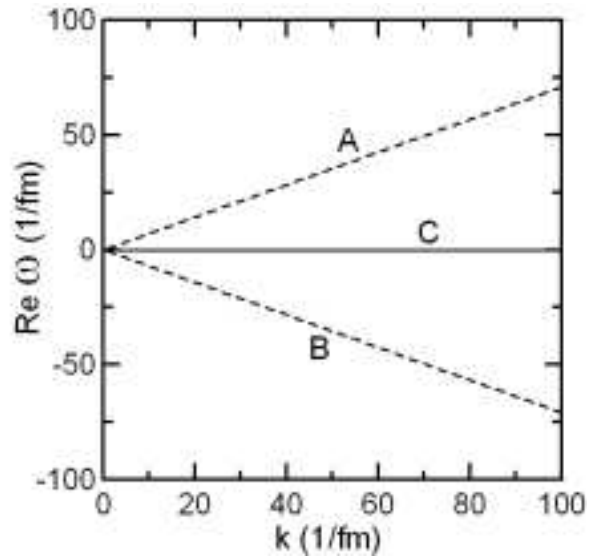}
\caption{The real part of the frequency $\protect\omega $ as a function of $%
k $. There are two propagating modes (dashed lines A and B) and
one non-propagating mode (solid line C).}
\label{fig1}
\end{figure}

To have the non-trivial solution for the perturbation, the determinant of
the $3\times 3$ matrix in the left-hand side of Eq.(\ref{perturb}) should
vanish so that $\omega $ should satisfy the following dispersion relation, 
\begin{equation}
\omega ^{2}-\alpha k^{2}=i\frac{\zeta }{(\epsilon +p)}\frac{\omega k^{2}}{%
1+i\tau _{R}\omega }.  \label{determinant}
\end{equation}%
whose solution for $\omega $ can be written as%
\begin{equation*}
\omega =x+\frac{A}{3}\frac{1}{x}+\frac{i}{3\tau _{R}},
\end{equation*}%
where 
\begin{eqnarray}
x &=&(-i)^{1/3}\sqrt{\frac{B}{2}+\sqrt{\left( \frac{B}{2}\right) ^{2}+\left( 
\frac{A}{3}\right) ^{3}}}, \\
A &=&\left( \frac{1}{b}+\alpha \right) k^{2}-\frac{1}{3\tau _{R}^{2}}, \\
B &=&\frac{1}{3\tau _{R}}\left( 2\alpha -\frac{1}{b}\right) k^{2}+\frac{2}{%
(3\tau _{R})^{3}},
\end{eqnarray}%
with 
\begin{equation}
\frac{1}{b}\equiv \frac{\zeta }{\tau _{R}}\frac{1}{\varepsilon +p}.
\label{gzi}
\end{equation}

\begin{figure}[tbp]
\includegraphics[scale=1]{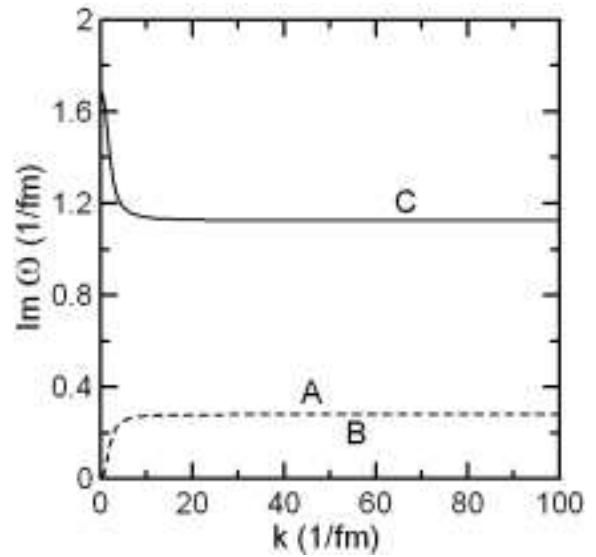}
\caption{The imaginary part of the frequency $\protect\omega $ as a function
of $k$. The two propagating modes A and B shown in Fig. \protect\ref{fig1}
are degenerated (dashed lines). The solid line is the non-propagating mode
C. }
\label{fig2}
\end{figure}

The asymptotic forms of the dispersion relation for large $k$ are 
\begin{equation}
\omega =\left\{ 
\begin{array}{c}
\pm k\sqrt{\frac{1}{b}+\alpha }+\frac{i}{2\tau _{R}\left( 1+\alpha b\right) }%
+O\left( k^{-1}\right) \\ 
i~\frac{\alpha b}{\tau _{R}\left( 1+\alpha b\right) }+O\left( k^{-1}\right)%
\end{array}%
\right. ,  \label{Disp}
\end{equation}%
whereas for small $k$, 
\begin{equation}
\omega =\left\{ 
\begin{array}{c}
\pm k\frac{\sqrt{\alpha -\left( \alpha +1/4b\right) \tau _{R}^{2}k^{2}/b}}{%
\left( 1-\tau _{R}^{2}k^{2}/b\right) }+\frac{ik^{2}\tau _{R}}{2\left( b-\tau
_{R}^{2}k^{2}\right) } \\ 
i/\tau _{R}~%
\end{array}%
\right. .  \label{Disp2}
\end{equation}%
We have three solutions; two of them are propagating modes and the remaining
one is a non-propagating mode ($\omega $ pure imaginary). From Eqs.(\ref%
{Disp}) and (\ref{Disp2}) we can see that the sound velocity of propagating
modes reduces to that of the ideal one for small $k$, on the other hand, for
large $k$, the sound velocity is given by $\sqrt{1/b+\alpha }$. In Figs. \ref%
{fig1} and \ref{fig2}, we show, respectively, the real and imaginary parts
of $\omega $ from these dispersion relations as functions of momentum $k$
for $a=0.1,$ $b=6,$ and the temperature $T=200\ MeV$. In Fig. \ref{fig1},
the lower two curves (solid and dot-dashed lines) correspond to the upper
line of Eq.(\ref{Disp}) and the horizontal line (dotted) corresponds to the
second line of Eq.(\ref{Disp}). As we see from Fig. \ref{fig2}, the
imaginary part of $\omega $ is always positive for all $k$ and converges to
constant values. Positivity of the imaginary part guarantees the stability
of the plane-wave perturbation.

We assume that the propagation of physical quantities for the propagating
modes are characterized by the group velocity. Then, from this relation, the
maximum velocity in such a theory is determined as 
\begin{equation}
v_{M}=\sqrt{1/b+\alpha }.  \label{phvel}
\end{equation}%
It should be noted that Eq. (\ref{phvel}) gives the velocity of sound in the
causal hydrodynamics. As a matter of fact, in the vanishing $\zeta $, this
definition coincides with the velocity of sound of ideal fluid. It is clear
that the group velocity becomes infinite at the limit of $\tau
_{R}\rightarrow 0$. In such a situation, always some portion of the matter
tries to propagate with velocity larger than the velocity of the light for
any initial condition. However, since the hydrodynamic equation is
covariant, the presence of the light-cone singularity forbids such a
propagation. Then the matter tends to accumulate at the light-cone and in
such situation the linearized wave analysis in the homogeneous back ground
at rest will breakdown. This kind of conflict between the causality and
relativistic covariance leads eventually to the instabilities of the
solution near the light-cone \cite{Hiscock,kouno,Gio}.

As for the non-propagating mode, we find that the imaginary part becomes a
constant for large $k.$ That is, the large $k$ component just damps
exponentially. This is different from the case of a diffusion process where
the non-propagating mode behaves as $k^{2}$ in the large $k$ limit, which
leads to an infinite propagation speed. In this sense, our theory is causal.

\subsection{Parameters}

There are so far two approaches to estimate the transport coefficients;
kinetic approach and microscopic approach. The calculation of the bulk
viscosity coefficients are much involved in the kinetic approach. We cannot
use the Boltzmann equation since it contains only the information of
two-body collisions. We have to use the (modified) Enskog model or the
Bogoliubov-Cho-Uhlenbeck equation where the multiple collision effects are
included. In the microscopic approach, it is known that the transport
coefficients are calculated by using the Green-Kubo-Nakano (GKN) formula.
However, the GKN formula is the formula for the Newtonian fluid like the
Navier-Stokes fluids, and hence we cannot use for the transport coefficients
of the causal dissipative hydrodynamics because it is a non-Newtonian fluid.
To calculate the transport coefficients of the causal dissipative
hydrodynamics, we have to derive a new formula. One possibility of such a
new formula is proposed in \cite{koide1,koide2,TKTK}. In any case, no
reliable estimate for the bulk viscosity coefficient is available.

In the present analysis, we will not deal with a precise quantitative
description of the behavior of the matter created in heavy-ion collisions
but rather interested in qualitative role of viscosity. Thus, we assume a
simple expression for the bulk viscosity coefficient $\zeta $ as usually
adopted for the shear viscosity, 
\begin{equation*}
\zeta =as,
\end{equation*}%
where $s$ is the entropy density in the local rest frame.

Another important parameter of the theory is the relaxation time $\tau _{R}.$
As was mentioned in the previous section, causality constraints the relation
between the two parameters, $\zeta $ and $\tau _{R}$ through $v_{M}\leq 1$
Thus, for the sake of simplicity, we parametrize the relaxation time by
taking $b$ as constant. This determines the relaxation time as 
\begin{equation*}
\tau _{R}=\frac{\zeta }{\epsilon +p}b.
\end{equation*}%
The later examples are presented in terms of these parameters $a$ and $b$.
However, there is no theoretical reason that $a$ and $b$ are constants, but
they may depend on thermodynamical quantities (see the later discussion). In
the following calculations, we use two values for the parameter $a$, $0.1$
and $1,$ with fixed value of $b=6$.

\section{Smoothed Particles Hydrodynamics}

To solve numerically the relativistic hydrodynamic equations we use the
Smoothed Particle Hydrodynamic (SPH) method. This method was initially
introduced \cite{SPH} for application in astrophysics. Using variational
approach, this method was extended to the application for heavy ion
collisions \cite{SPH-heavy ion}.

The original idea of the SPH method is to obtain an approximate solution of
hydrodynamics by parameterizing the fluid into a set of ``effective
particles\textquotedblright . However, in the application to the heavy ion
dynamics, the SPH method is not a mere mathematical discretization scheme,
but can be interpreted as a physical model of the collective motion in terms
of a finite set of dynamical variables.

To see this, let us consider a distribution $a\left( \mathbf{r},t\right)$ of
any extensive physical quantity, $A$. In a system like the hot and dense
matter created in heavy ion collisions, the behavior of $a(\mathbf{r},t)$
contains the effects of whole microscopic degrees of freedom. We are not
interested in the extremely short wavelength behavior of $a\left( \mathbf{r}%
,t\right)$ but rather in global behaviors which are related directly to the
experimental observables. Therefore, we would like to introduce a
coarse-graining procedure for $a$. To do this, we introduce the kernel
function $W\left( \mathbf{r}-\mathbf{\tilde{r}},h\right)$ which maps the
original distribution $a$ to a coarse-grained version $a_{CG}$ as, 
\begin{equation}
a_{CG}\left( \mathbf{r},t\right) =\int a(\mathbf{\tilde{r}},t)W\left( 
\mathbf{r}-\mathbf{\tilde{r}},h\right) d\mathbf{\tilde{r}}  \label{as}
\end{equation}%
where $W$ is normalized, 
\begin{equation}
\int W\left( \mathbf{\tilde{r}},h\right) d\mathbf{\tilde{r}=1,}
\label{normalW}
\end{equation}%
and has a bounded support of the scale of $h$, 
\begin{equation*}
W\left( \mathbf{r},h\right) \rightarrow 0,\ \ \left\vert \mathbf{r}%
\right\vert \gtrsim h,
\end{equation*}
satisfying 
\begin{equation*}
\underset{h\rightarrow 0}{\lim }W\left( \mathbf{\tilde{r}},h\right) =\delta
\left( \mathbf{\tilde{r}}\right) .
\end{equation*}%
Here, $h$ is a typical length scale for the coarse-graining in the sense
that the kernel function $W$ introduces a cut-off in short wavelength of the
order of $h$. Thus we will take this value as the scale of coarse graining
in the QCD dynamics (i.e., the mean-free path of partons) to obtain the
hydrodynamics of QGP ($h\simeq 0.1fm$).

The second step is to approximate this coarse grained distribution $a_{CG}(%
\mathbf{r},t)$ by replacing the integral in Eq.(\ref{as}) by a summation
over a finite and discrete set of points, $\left\{ \mathbf{r}_{\alpha
}(t),\alpha =1,..,N_{SPH}\right\} ,$%
\begin{equation}
a_{SPH}\left( \mathbf{r},t\right) =\sum_{\alpha =1}^{N_{SPH}}A_{\alpha
}\left( t\right) W(|\mathbf{r}-\mathbf{r}_{\alpha }(t)|).  \label{aSPH}
\end{equation}%
If the choice of $\left\{ A_{\alpha }(t),\alpha =1,..,N_{SPH}\right\} $ and $%
\left\{ \mathbf{r}_{\alpha }(t),\alpha =1,..,N_{SPH}\right\} $ are
appropriate, the above expression should converge to the coarse-grained
distribution $a_{CG}$ for large $N_{SPH}.$ Parameters $\left\{ A_{\alpha
}(t),\alpha =1,..,N_{SPH}\right\} $ and $\left\{ \mathbf{r}_{\alpha
}(t),\alpha =1,..,N_{SPH}\right\} $ should be determined from the dynamics
of the system. In practice, we first choose the reference density $\sigma
^{\ast }$ which is conserved,%
\begin{equation}
\frac{\partial \sigma ^{\ast }}{\partial t}+\nabla \cdot \mathbf{j}=0,
\label{conti}
\end{equation}%
where $\vec{j}$ is the current associated with the density $\sigma ^{\ast }.$
Then, we note that the following ansatzs, 
\begin{eqnarray*}
\sigma _{SPH}^{\ast }\left( \mathbf{r},t\right) &=&\sum_{\alpha
=1}^{N_{SPH}}\nu _{\alpha }W(|\mathbf{r}-\mathbf{r}_{\alpha }(t)|), \\
\mathbf{j}_{SPH}\left( \mathbf{r},t\right) &=&\sum_{\alpha =1}^{N_{SPH}}\nu
_{\alpha }\frac{d\mathbf{r}_{\alpha }(t)}{dt}W(|\mathbf{r}-\mathbf{r}%
_{\alpha }(t)|),
\end{eqnarray*}%
satisfies the equation, 
\begin{equation*}
\frac{\partial \sigma _{SPH}^{\ast }}{\partial t}+\nabla \cdot \mathbf{j}%
_{SPH}=0,
\end{equation*}
where $\nu _{\alpha} 
{\acute{}}%
s$ are constant. By using the normalization of $W,$ Eq.(\ref{normalW}), we
have 
\begin{equation*}
\int_{SPH}\sigma ^{\ast }\left( \mathbf{r},t\right) d^{3}\mathbf{r=}%
\sum_{\alpha =1}^{N_{SPH}}\nu _{\alpha }.
\end{equation*}%
Then we can interpret the quantity $\nu _{\alpha }$ as the conserved
quantity attached at the point $\mathbf{r}=\mathbf{r}_{\alpha }(t)$.
Therefore, the distribution $\sigma _{SPH}^{\ast }\left( \mathbf{r},t\right) 
$ is a sum of small piece-wise distribution, carrying the density,%
\begin{equation*}
\nu _{\alpha }W(|\mathbf{r}-\mathbf{r}_{\alpha }(t)|).
\end{equation*}%
These pieces are referred to as "SPH-particles".

Using the above reference density and the extensive nature of $A,$ we can
write $A_{\alpha }$ in Eq.(\ref{aSPH}) as%
\begin{equation*}
A_{\alpha }\left( t\right) =\nu _{\alpha }\frac{a(\mathbf{r}_{\alpha },t)}{%
\sigma ^{\ast }(\mathbf{r}_{\alpha },t)}
\end{equation*}%
which represents the quantity $A$ carried by the SPH particle at the
position $\mathbf{r}=\mathbf{r}_{\alpha }(t)$. In fact, the total amount of $%
A$ of the system at the instant $t$ is given by%
\begin{equation*}
A\left( t\right) =\sum_{\alpha =1}^{N_{SPH}}A_{\alpha }\left( t\right) .
\end{equation*}%
In the previous works the entropy density is chosen as the reference density
and the dynamics of the parameters $\left\{ \mathbf{r}_{\alpha }(t),\alpha
=1,..,N_{SPH}\right\} $ are determined from the variational principle from
the action of ideal hydrodynamics \cite{SPH-heavy ion}. Thus, the SPH
particle coordinates and their time derivatives are considered as the
variational parameters which optimize dynamically the action of the system.

The entropy density is, however, conserved only for the motion of an ideal
fluid. When we discuss the behavior of viscous fluids, where the entropy is
not conserved, we cannot use the entropy density as the reference SPH
density. Thus we introduce a new conserved quantity, the specific proper
density $\sigma $, which is defined by the flow of the fluid, 
\begin{equation}
\partial _{\mu }\left( \sigma u^{\mu }\right) =0,  \label{conserv-sigmav}
\end{equation}%
and we will use it as the reference density for viscous fluids. Here, the
four-velocity $u^{\mu }$ is defined in terms of the local rest frame of the
energy flow (Landau frame). The specific density is expressed in the SPH
form as 
\begin{equation*}
\sigma ^{\ast }(\mathbf{r},t)=\sum_{\alpha =1}^{N_{SPH}}\nu _{\alpha }W(|%
\mathbf{r}-\mathbf{r}_{\alpha }(t)|),
\end{equation*}%
where $\sigma ^{\ast }=\sigma u^{0}$ is the specific density in the
laboratory frame and $\nu _{\alpha }$ is the inverse of the specific volume
of the SPH particle $\alpha ,$ and is chosen as an arbitrary constant. Final
results do not depend on this choice and we set $\nu _{\alpha }=1$ for
simplicity. As for the kernel $W(\mathbf{r})$, we use the spline function 
\cite{SPH}.

Strictly speaking, this procedure is only possible provided that the lines
of flow in space defined by the velocity field $u^{\mu }$ do not cross each
other during the evolution in time. That is, if there appear turbulences or
singularities in the flow lines, the above definition of Lagrange
coordinates fails. However, if the size of $h$ is consistently chosen as the
size of coarse graining of the underlying microscopic theory, the flux line
calculated using this $h$ should not cross.

Now we apply this method to the causal dissipative hydrodynamics in $1+1$
dimension. We have to solve the evolution equation of the viscosity in the
SPH scheme. To do so, we express the viscosity as 
\begin{equation}
\Pi =\sum_{\alpha =1}^{N_{SPH}}\nu _{\alpha }\frac{\Pi _{\alpha }}{\sigma
_{\alpha }^{\ast }}W(|\mathbf{r}-\mathbf{r}_{\alpha }(t)|),
\label{Binterpolation}
\end{equation}%
Time evolution of the term $\Pi _{\alpha }$ can be calculated as%
\begin{equation}
\gamma _{\alpha }\frac{d\Pi _{\alpha }}{dt}=-\frac{\zeta }{\tau _{R}}\left(
\partial _{\mu }u^{\mu }\right) _{\alpha }-\frac{1}{\tau _{R}}\Pi _{\alpha }
\label{BulkSPH}
\end{equation}%
where $\gamma _{\alpha }$ is the Lorentz factor of the $\alpha -th$
particle. At the same time, using the SPH expression for the entropy density 
$s^{\ast }$ in the observable frame, 
\begin{equation}
s^{\ast }=\sum_{\alpha =1}^{N_{SPH}}\nu _{\alpha }\left( \frac{s}{\sigma }%
\right) _{\alpha }W(|\mathbf{r}-\mathbf{r}_{\alpha }(t)|),
\label{Sinterpolation}
\end{equation}%
and using Eq.(\ref{s-current})\ we find,%
\begin{equation*}
\frac{d}{dt}\left( \frac{s}{\sigma }\right) _{\alpha }=-\frac{1}{T}\frac{\Pi
_{\alpha }}{\sigma _{\alpha }^{\ast }}\left( \partial _{\mu }u^{\mu }\right)
_{\alpha }.
\end{equation*}%
where $s=s^{\ast }/u^{0}$ is the proper entropy density. In the following,
we denote the quantity in the observable frame with the asterisk. In the
above expressions, relaxation time $\tau _{R},$ viscosity coefficient $\zeta 
$ and temperature $T$ are functions of space and time, so that they should
be evaluated at the position of each particle $\alpha .$

Finally, we need to express the momentum conservation equation by the SPH
variables. We write the space component of Eq.(\ref{divTmunu}) in terms of
the reference density, 
\begin{equation}
\sigma \frac{d}{d\tau }\left( \frac{\epsilon +p+\Pi }{\sigma }u^{i}\right)
+\partial _{i}(p+\Pi )=0.  \label{Motion}
\end{equation}

It should be noted that there exist ambiguities within the resolution of the
coarse-graining size $h$ to express the equation of motion in the SPH form.
However, in the ideal fluid, the SPH equation of motion can be derived by
the variational method uniquely. Thus, we obtain the equation of motion by
using the same SPH parametrization to Eq.(\ref{Motion})\textbf{, }

\begin{gather}
\sigma _{\alpha }\frac{d}{d\tau _{\alpha }}\left( \frac{\epsilon _{\alpha
}+p_{\alpha }+\Pi _{\alpha }}{\sigma _{\alpha }}u_{\alpha }^{i}\right) = 
\notag \\
\sum_{\beta =1}^{N_{SPH}}\nu _{\beta }\sigma _{\alpha }^{\ast }\left( \frac{%
p_{\beta }+\Pi _{\beta }}{\left( \sigma _{\beta }^{\ast }\right) ^{2}}+\frac{%
p_{\alpha }+\Pi _{\alpha }}{\left( \sigma _{\alpha }^{\ast }\right) ^{2}}%
\right) \partial _{i}W(|\mathbf{r}_{\alpha }-\mathbf{r}_{\beta }(t)|),
\label{MotionSPH}
\end{gather}%
where the right hand side of Eq.(\ref{MotionSPH}) corresponds to the term $%
\partial _{i}(p+\Pi )$ written in terms of the SPH parametrization. We
remark that in the case of vanishing viscosity our result is reduced to the
expression derived with variational principle for ideal fluids.

By separating the acceleration and force terms in Eq.(\ref{MotionSPH}), we
obtain our final expression of the equation of motion for the SPH particles,

\begin{equation*}
M_{\alpha }\frac{d\vec{u}_{\alpha }}{dt}=\vec{F}_{\alpha },
\end{equation*}%
where 
\begin{eqnarray}
M_{ij} &=&\gamma (\epsilon +p+\Pi )\delta _{ij}+Au_{i}u_{j},  \label{Mij} \\
F_{j} &=&-\partial _{j}\left( p+\Pi \right) +Bu_{j},  \label{Fj}
\end{eqnarray}%
with 
\begin{eqnarray}
A &=&\frac{1}{\gamma }\left[ \varepsilon +p+\Pi -\frac{\partial }{\partial s}%
\left( \varepsilon +p\right) \left( s+\frac{\Pi }{T}\right) -\frac{\zeta }{%
\tau _{R}}\right] ,  \label{A} \\
B &=&A\frac{\gamma ^{2}}{\sigma ^{\ast }}\frac{d\sigma ^{\ast }}{dt}+\frac{%
\Pi }{\tau _{R}}.  \label{B}
\end{eqnarray}%
These set of equations define the coarse-grained dynamics to represent the
continuity equation for the energy and momentum tensor of a relativistic
fluid, together with an irreversible mechanism which converts a part of
collective kinetic energy (the motion of SPH particles) into internal heat
of the fluid.

For the sake of book-keeping, we summarize the practical algorithm of
calculating the SPH method to solve numerically the causal dissipative
hydrodynamics. The dynamics is described by the following variables
corresponding to the quantities attached to the each SPH particle:%
\begin{equation*}
\left\{ \mathbf{r}_{\alpha },\mathbf{u}_{\alpha },\nu _{\alpha },\left( 
\frac{s}{\sigma }\right) _{\alpha },\Pi _{\alpha };\ \alpha
=1,..,N_{SPH}\right\} .
\end{equation*}%
At the initial time, their values are determined according to the initial
condition. The entropy density profile and the bulk viscosity are then
obtained with the interpolations in Eq.(\ref{Sinterpolation}) and Eq.(\ref%
{Binterpolation}), respectively. The energy density, pressure and
temperature are calculated with the equation of state. The time evolution of
these quantities are calculated by solving the equations derived previously,%
\begin{eqnarray*}
\frac{d\vec{u}_{\alpha }}{dt} &=&M_{\alpha }^{-1}\vec{F}_{\alpha }, \\
\gamma _{\alpha }\frac{d\Pi _{\alpha }}{dt} &=&-\frac{\zeta }{\tau _{R}}%
\left( \partial _{\mu }u^{\mu }\right) _{\alpha }-\frac{1}{\tau _{R}}\Pi
_{\alpha }, \\
\frac{d}{dt}\left( \frac{s}{\sigma }\right) _{\alpha } &=&-\frac{1}{T}\frac{%
\Pi _{\alpha }}{\sigma _{\alpha }^{\ast }}\left( \partial _{\mu }u^{\mu
}\right) _{\alpha }
\end{eqnarray*}%
Here the inverse matrix of $M$ is calculated explicitly as%
\begin{eqnarray*}
M^{-1} &=&\frac{1}{\gamma (\epsilon +p+\Pi )}\widehat{1} \\
&&-\frac{A}{\gamma (\epsilon +p+\Pi )\left( \gamma (\epsilon +p+\Pi )+\left(
\gamma ^{2}-1\right) A\right) }\vec{u}\ \vec{u}^{T}.
\end{eqnarray*}%
and the four-divergence of the velocity is calculated as%
\begin{equation*}
\left( \partial _{\mu }u^{\mu }\right) _{a}=-\frac{\gamma _{\alpha }}{\sigma
_{\alpha }^{\ast }}\left( \frac{d\sigma ^{\ast }}{dt}\right) _{\alpha }+%
\frac{1}{\gamma _{\alpha }}\vec{u}_{\alpha }\cdot \frac{d\vec{u}_{\alpha }}{%
dt}
\end{equation*}%
with 
\begin{equation*}
\frac{d\sigma ^{\ast }}{dt}=\frac{1}{\sigma ^{\ast }}\ \vec{j}\cdot \nabla
\sigma ^{\ast }-\nabla \cdot \vec{j}.
\end{equation*}

\section{Examples}

\subsection{Expansion to the vacuum and stationary boundary}

Let us consider first the Landau type initial condition. For the ideal case,
this example has already been discussed in \cite{SPHERIO} and the SPH scheme
works very well. When we introduce the viscosity, we found that the sharp
discontinuity in the boundary leads to undesirable instabilities. As we will
discuss later, the origin of these instabilities is due to the presence of a
space discontinuity. When we relax such a steep boundary by replacing the
initial distribution consistent with the SPH scale used, the above
instabilities disappears. In this subsection, we use basically the Landau
type initial distribution with the temperature $200\ MeV$, distributed
uniformly within the range $x\in \left[ -1,1\right] \ fm.$ To relax the
sharp boundaries, we add the surface thickness of $10h,$ where $h=0.01$ $fm$%
. For simplicity, we take both vanishing initial velocity and viscosity, 
\begin{equation*}
u\left( x,0\right) =0,\ \Pi (x,0)=0.
\end{equation*}%
As we mentioned in the introduction, we use the equation of state that of
massless ideal gas, 
\begin{equation}
p=\frac{\epsilon }{3},
\end{equation}%
which in turn, 
\begin{equation*}
\epsilon =Cs^{4/3},\ \ T=\frac{4}{3}Cs^{1/3},
\end{equation*}%
where $C$ is a constant related to the Stephan-Boltzmann constant of
massless $3$ flavor quark-gluon gas.

\begin{figure}[tbp]
\includegraphics[scale=1]{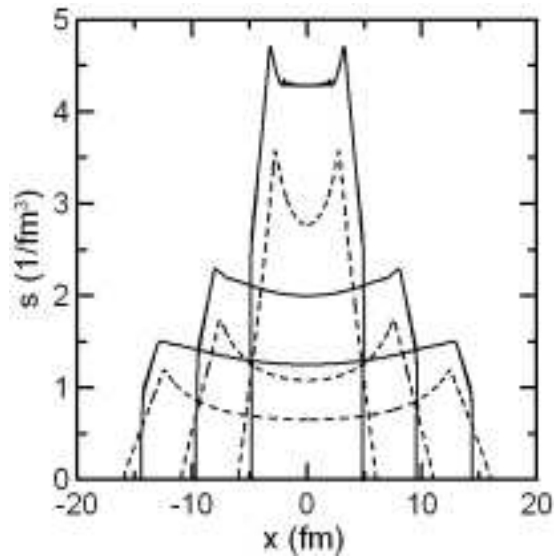}
\caption{The time evolutions of the proper entropy density for ideal fluid
(dotted line) and the viscous fluid of $a=0.1$ (solid line).}
\label{fig3}
\end{figure}

The temporal evolution of the density profile for the entropy at $t=5,10$
and $15$ fm is shown in Fig.\ref{fig3}. The solid line represents the
results of the causal hydrodynamics with small viscosity, $a=0.1$ and $b=6$.
For the sake of comparison, we show the time evolution of the same initial
condition for the ideal fluid with the dashed lines.

For the massless ideal fluid, we know that the propagation speed into the
vacuum should be the speed of light. On the other hand, one can see that the
propagation speed into the vacuum for the viscous fluid is slower than that
of the ideal fluid. This is because the viscosity acts as an attractive
interaction during the expansion of the fluid. Thus it takes more time to
achieve the speed of light in the causal dissipative hydrodynamics. On the
other hand, we can also observe that the propagation of rarefaction wave
into the matter of the viscous fluid is faster than that of the ideal case.
This is what we expect from Eq.(\ref{phvel}).

\begin{figure}[tbp]
\includegraphics[scale=1]{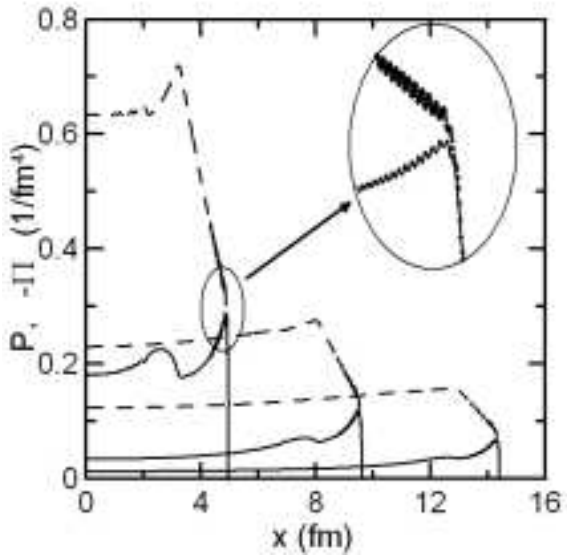}
\caption{The time evolution of the pressure $P$ (dotted) and viscosity $\Pi $
(solid) of the fluid. At the boundary, we can see the relation $P=-\Pi $.
The in-set shows the details of the region where the two curves starts to
coincide.}
\label{fig4}
\end{figure}

Interestingly enough, we observe that the behavior of the boundary of the
viscous fluid seems to be a kind of a stationary wave. To understand this
phenomenon, let us introduce the relative coordinate $y=x-v_{s}t$ with a
velocity of the stationary wave $v_{s}$ and suppose that the energy-momentum
tensor depends only on $y$ near the boundary. Then, from the equation of
continuity of the energy-momentum tensor, we have 
\begin{eqnarray}
\frac{d}{dy}[-v_{s}T^{00}+T^{01}] &=&0, \\
\frac{d}{dy}[-v_{s}T^{01}+T^{11}] &=&0.
\end{eqnarray}%
where, 
\begin{equation*}
T^{\mu \nu }=\left( \varepsilon +P+\Pi \right) u^{\mu }u^{\nu }-g^{\mu \nu
}\left( p+\Pi \right) .
\end{equation*}%
and $u^{\mu }=\left( \gamma ,\gamma v\right) .$ From the boundary condition
at the vacuum where $T^{\mu \nu }$ vanishes, we get the following two
conditions to define the stationary wave, 
\begin{eqnarray}
v_{s}(T^{11}+T^{00}) &=&T^{01}(1+v_{s}^{2}),  \label{eqn:cd1} \\
T^{11} &=&v_{s}^{2}T^{00}.  \label{eqn:cd2}
\end{eqnarray}%
From Eq. (\ref{eqn:cd1}), we find $v_{s}=v$ or $1/v$. The first solution is
consistent of our stationary wave assumption, showing that the fluid
propagates to the vacuum with the velocity $v$. The second solution leads to
a trivial situation $\epsilon =P=0$. When we substitute $v_{s}=v$ into Eq.(%
\ref{eqn:cd2}) we get $P=-\Pi $. Thus, if the boundary of the fluid
propagates as a stationary wave, then the pressure and the viscosity should
satisfy this relation. In fact, this relation corresponds to the case where
the acceleration vanishes for the fluid motion. In Fig.\ref{fig4}, we
plotted the profiles of $P$ and $\Pi $ and we can see clearly the relation $%
P=-\Pi $ is satisfied near the boundary (see the in-set). In this figure, we
notice that there appear quick oscillation modes with the wavelength of the
order of few $h.$ We will discuss this point in the following sections.

We have checked for various combinations of parameters and we found that the
above feature of the appearance of a stationary wave near the boundary is
universal for the free expansion of viscous matter into the vacuum. This
fact is very important because the viscosity $\Pi $ is always the same order
as the pressure $P$ near the boundary to the vacuum, showing that the
viscous effect can not be treated as a small correction to the equilibrium
thermodynamical quantities.

\subsection{Density discontinuity and shock wave}

\begin{figure}[tbp]
\includegraphics[scale=1]{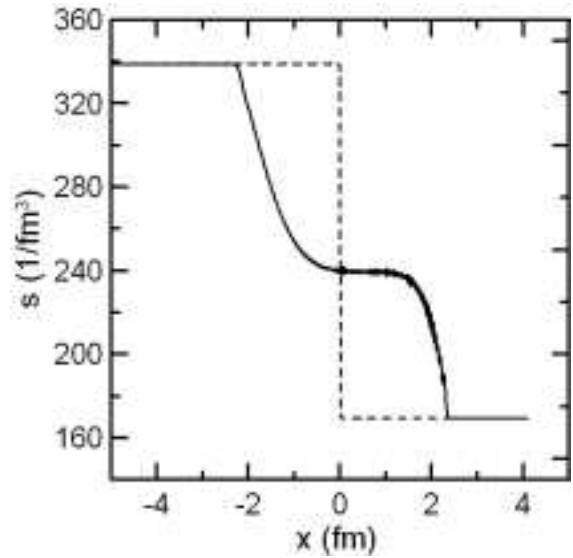}
\caption{The shock formation in the proper entropy profile (solid) of the
viscous fluid of $a=0.1$, starting from the discontinuous initial condition
(dotted). }
\label{fig5}
\end{figure}

\begin{figure}[tbp]
\includegraphics[scale=1]{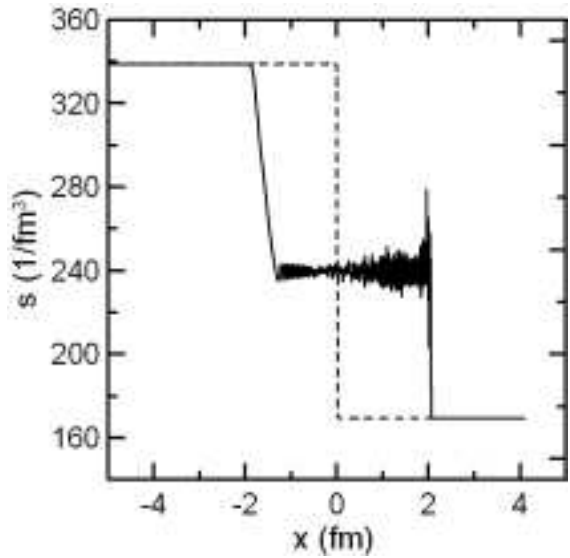}
\caption{The behavior of SPH solution for the ideal fluid in the presence of
shock wave corresponding to Fig. \protect\ref{fig5}.}
\label{fig6}
\end{figure}

One interesting question is whether the dissipative hydrodynamics can
describe the formation of a shock wave. We know that for an ideal fluid,
when a shock is formed, we need the so-called artificial viscosity for
smoothing the shock region. One might argue that when the real viscosity is
present, we do not need to introduce such an artificial viscosity. In Fig. %
\ref{fig5}, we show the time evolution of the entropy density of a viscous
fluid of $a=0.1$\ whose initial distribution has a discontinuity. We know
that for the case of an ideal fluid, such an initial distribution leads to a
shock wave propagation into the medium. The thickness of the shock front
should be zero for an ideal fluid. With the presence of a finite viscosity,
the shock wave is smoothed and appears just a quick change in density. For
the comparison, we show in Fig. \ref{fig6} the time evolution of the density
profile for an ideal case calculated by the SPH scheme with a finite $h$,
where we see a quick oscillation of large amplitude. It is interesting to
observe that the wavelength of these oscillations is exactly the order of $h$
of the SPH scheme.

The physical reason for these oscillations of the SPH particles is clear.
From the well-known Hugoniot-Ranking relation, we know that at the shock
front, there should exist a production of entropy \cite{LL}. However, for an
ideal fluid, the total entropy is conserved so that the SPH particles carry
the extra kinetic energy corresponding to the entropy production at the
shock. These extra oscillations propagate with the smallest wavelength
(order of $h$ ). The presence of viscosity can damp these oscillations and
turns the kinetic energy of the SPH particles to internal heat of the fluid,
recovering the Hugoniot-Ranking relation. This is the case of the example in
Fig. \ref{fig5}.

\begin{figure}[tbp]
\includegraphics[scale=1]{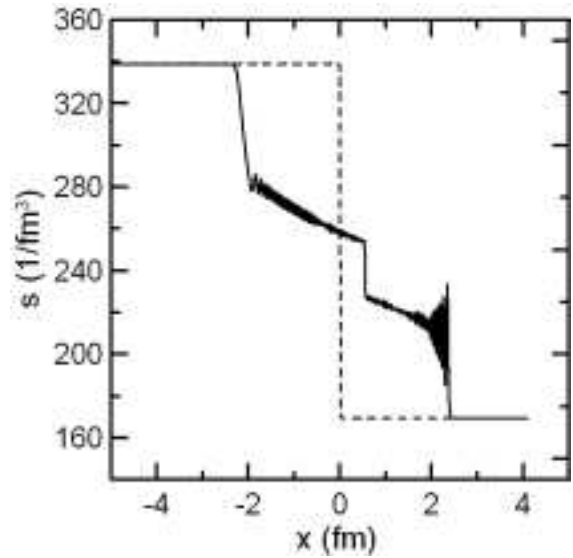}
\caption{The shock formation in the proper entropy profile (solid) of the
viscous fluid of $a=1$, starting from the discontinuous initial condition
(dotted). }
\label{fig7}
\end{figure}

However, this is not always the case. For example, when $a$ becomes large,
the relaxation time $\tau _{R}$ increases, because of causality. Then the
time scale for the damping becomes comparable to the evolution time scale
generating an oscillatory behavior. Such example is shown in Fig. \ref{fig7}
with $a=1$. In addition, when we look precisely the density profile shown in
the example of Fig. \ref{fig5}, we note the existence of quick oscillation
modes with the wavelength of the order of $h,$\ although in this example,
their amplitude does not increase in time.

In the SPH calculation,\ by assumption, $h$\ is the scale of coarse graining
and the resulting dynamics should always have larger wavelength than this.
Thus, it is clear that the appearance of such rapidly oscillating modes of
short wavelength of the order of $h$\ is a signature of an inconsistency of
the theory with this finite coarse-graining scale.\textbf{\ }Such a
situation also happens in a simple expansion of the fluid into the vacuum in
the presence of viscosity. As mentioned in the beginning of this section, if
we take a real sharp discontinuous Landau initial condition, similar rapidly
oscillating modes appear at the expanding boundary. This is because a real
discontinuity is not compatible with a finite coarse-graining scale $h.$ See
also the small oscillations in Fig. \ref{fig4}. When we discuss the shock
phenomena using an ideal fluid, then the shock front has null thickness and
it should be treated as a singularity of the theory. This is because the
usual hydrodynamics assume a null coarse-graining scale.

Note that the usual viscosity is also obtained assuming a vanishing
coarse-graining scale. To be consistent with the coarse-graining scheme of
the theory, there should be an additional dissipation mechanism which is
related to the coarse-graining scale. Such a mechanism should vanish in the
limit of $h/L\ll 1,$ where $L$ is the typical observable scale of the
dynamics.

\section{Viscosity associated with coarse graining scale}

As discussed above, we need an additional viscosity, whose scale is
determined by $h.$ Such a viscosity also should obey the requirement of
causality. Thus we propose%
\begin{equation*}
\Pi _{Tot}=\Pi +\Pi _{h},
\end{equation*}%
where as before%
\begin{equation*}
\tau _{R}^{\left( h\right) }\frac{d\Pi _{h}}{d\tau }=-\Pi _{h}-\zeta
^{\left( h\right) }\partial _{\mu }u^{\mu }
\end{equation*}%
with%
\begin{eqnarray*}
\zeta ^{\left( h\right) } &=&a^{\left( h\right) }\left( \varepsilon
+P\right) h, \\
\tau _{R}^{\left( h\right) } &=&\frac{\zeta ^{\left( h\right) }}{\epsilon +p}%
b^{\left( h\right) }.
\end{eqnarray*}%
The coefficients $a^{\left( h\right) }$ and $b^{\left( h\right) }$ are the
numbers of the order of unity. Strictly speaking, these numbers should be
determined from a microscopic theory by incorporating the effect of the
coarse-graining scale appropriately. Here we take them as phenomenological
parameters and found that the values,%
\begin{equation}
a^{\left( h\right) }=1/2,\ \ \ b^{\left( h\right) }=2,  \label{ah&bh}
\end{equation}%
can eliminate undesirable oscillations in the SPH motion.

When this additional viscosity is included, the viscosity term $\Pi $ in
Eqs.(\ref{Mij}) and (\ref{Fj}) should be replaced by $\Pi _{Tot}$ and the
coefficients $A$ and $B$ in Eqs.(\ref{A}) and (\ref{B}) are modified as 
\begin{align}
A& \rightarrow \frac{1}{\gamma }\left[ \varepsilon +p+\Pi _{Tot}-\frac{%
\partial \left( \varepsilon +p\right) }{\partial s}\left( s+\frac{\Pi _{Tot}%
}{T}\right) -\frac{\zeta }{\tau _{R}}-\frac{\zeta ^{\left( h\right) }}{\tau
_{R}^{\left( h\right) }}\right] ,  \notag \\
& \\
B& =A\frac{\gamma ^{2}}{\sigma ^{\ast }}\frac{d\sigma ^{\ast }}{dt}+\frac{%
\Pi }{\tau _{R}}+\frac{\Pi ^{\left( h\right) }}{\tau _{R}^{\left( h\right) }}%
.
\end{align}%
Consequently, the dispersion relation for the linearized sound wave, Eq.(\ref%
{determinant}) is modified as%
\begin{equation}
\omega ^{2}-\alpha k^{2}=i\frac{\zeta }{(\epsilon +p)}\frac{\omega k^{2}}{%
1+i\tau _{R}\omega }+i\frac{\zeta ^{\left( h\right) }}{(\epsilon +p)}\frac{%
\omega k^{2}}{1+i\tau _{R}^{\left( h\right) }\omega }.  \label{4th order}
\end{equation}

\begin{figure}[tbp]
\includegraphics[scale=1]{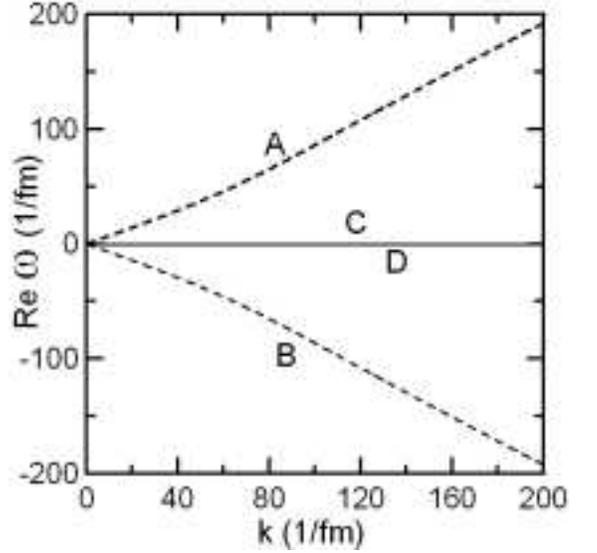}
\caption{The real part of the frequency $\protect\omega $ as a function of $%
k $. There are two propagating modes A, B (dashed lines) and two
non-propagating modes C and D which are degenerated (solid). }
\label{fig8}
\end{figure}

\begin{figure}[tbp]
\includegraphics[scale=1]{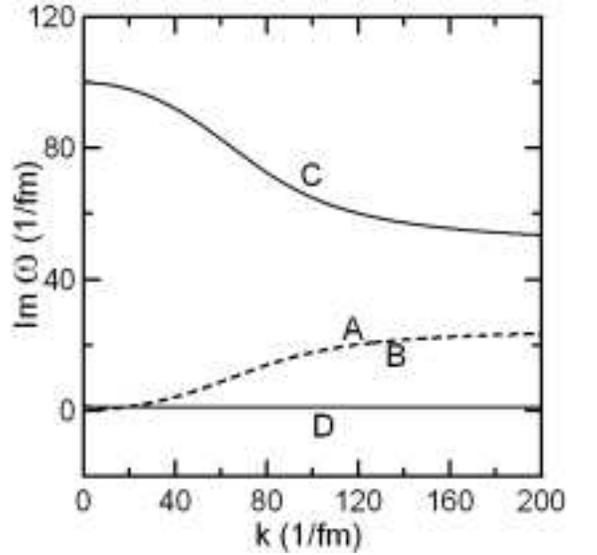}
\caption{The imaginary part of the frequency $\protect\omega $ as a function
of $k$. The two propagating modes A and B are degenrated (dashed line). The
non-propagating modes C and D have different $k$ dependence. }
\label{fig9}
\end{figure}

In Figs. \ref{fig8} and \ref{fig9}, we show the real and imaginary parts,
respectively, of the frequency $\omega $ as function of $k$, calculated from
Eq.(\ref{4th order}) for $a=0.1,$ $b=6$ and $T=200\ MeV$. This time, there
are 4 solutions, but only 2 of them are propagating modes and the other 2
are non-propagating modes. All them have positive imaginary part so that our
equation is stable to a linear perturbation around a hydrostatic state.

Under the normal condition, $\tau _{R}^{\left( h\right) }/\tau _{R}\ll 1,$\ $%
\zeta ^{\left( h\right) }\ll \zeta ,$ the above equation gives the same
dispersion relation as the previous case, as far as $\tau _{R}^{\left(
h\right) }\omega \,<1,$ since the second term of the right-hand side of Eq.(%
\ref{4th order}) can be neglected compared to the first term. On the other
hand, for the very large $k$ limit, the asymptotic form of the dispersion
relation becomes%
\begin{equation*}
\omega \rightarrow \pm \sqrt{\alpha +\frac{1}{b}+\frac{1}{b^{\left( h\right)
}}}k+i\frac{1}{2}\frac{1/(b\tau _{R})+1/(b^{\left( h\right) }\tau
_{R}^{\left( h\right) })}{\alpha +1/b+1/b^{\left( h\right) }}.
\end{equation*}%
The most effective choice of the additional viscosity is obtained for the
smallest possible value of $\tau _{R}^{\left( h\right) },$ preserving
causality. This determines the value of $b^{\left( h\right) }$ as%
\begin{equation*}
b^{\left( h\right) }=\frac{1}{1-\alpha -1/b}.
\end{equation*}%
The choice Eq.(\ref{ah&bh}) corresponds to the case $b=6$ and $\alpha =1/3.$
The difference of the two regimes determined by each viscosity can be seen
in these figures.

The additional viscosity introduced here can be consistent with the
stability and causality of the theory. With this procedure, the rapid
oscillation modes associated with the degrees of freedom beyond the
applicability of the coarse-grained theory can be naturally eliminated.

\subsection{\textquotedblleft Double Shock\textquotedblright\ Phenomena}

\begin{figure}[tbp]
\includegraphics[scale=1]{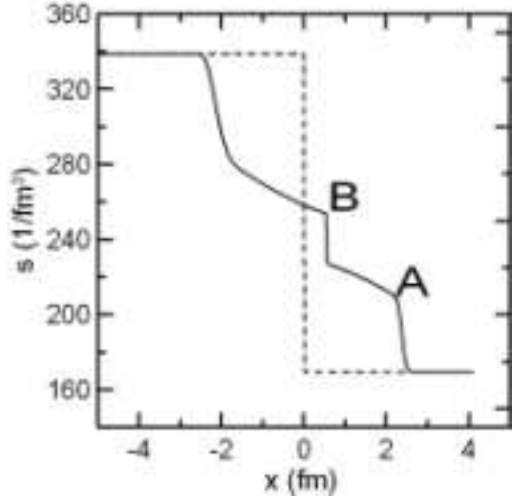}
\caption{The same example as Fig. \protect\ref{fig7}, calculated with the
additional viscosity. Quick osculations at the shock region are removed.}
\label{fig10}
\end{figure}

We have already shown that, without the additional viscosity, the SPH time
evolution develops rapidly oscillating modes with large amplitude whose
wavelength are of the order of $h.$ In Fig. \ref{fig10}, we show the result
of the same time evolution with the additional viscosity. We see that this
new viscosity successfully smears out the quick oscillating modes.

A very interesting feature of this example, in Fig. \ref{fig10}, is the
appearance of non-trivial structure of discontinuities when compared to the
usual shock discontinuity seen in Fig. \ref{fig4}. By changing the viscosity
parameter $a$ from the small value to the large ones, we found that the
discontinuity associated with the usual shock corresponds to the
discontinuity A shown in this figure. The new discontinuity B starts to
appear at some critical value of parameter $a.$ This can be interpreted as
the transportation of the discontinuity in the initial distribution by a
causal diffusion mechanism, as is known in the case of matter transportation
by a telegraph equation (See Appendix ). In fact, the velocity profile for
this apparent \textquotedblleft double shock\textquotedblright\ phenomena
does not have any discontinuity at the location of B as is shown in Fig. \ref%
{fig11}.

\begin{figure}[tbp]
\includegraphics[scale=1]{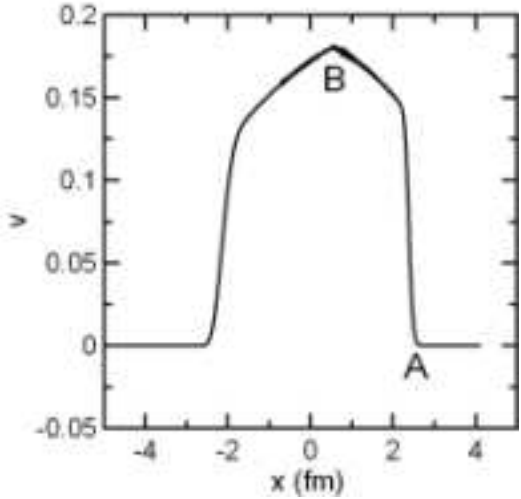}
\caption{The velocity profile corresponding to Fig. \protect\ref{fig10}.}
\label{fig11}
\end{figure}

\subsection{Shock wave Formation by quickly moving fluid component}

\begin{figure}[tbp]
\includegraphics[scale=1]{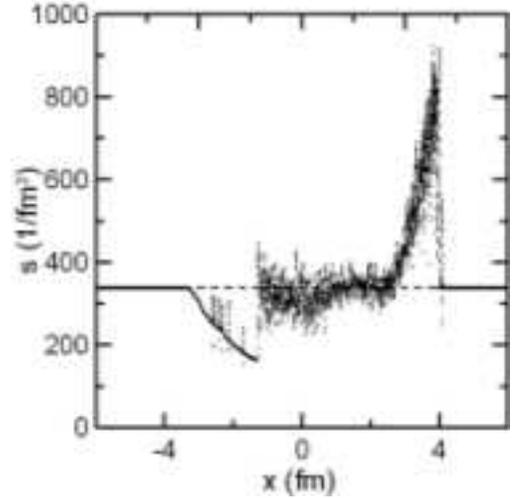}
\caption{The shock formation of the proper entropy density of the ideal
fluid, starting from the homogeneous initial condition (dotted). }
\label{fig12a}
\end{figure}

\begin{figure}[tbp]
\includegraphics[scale=1]{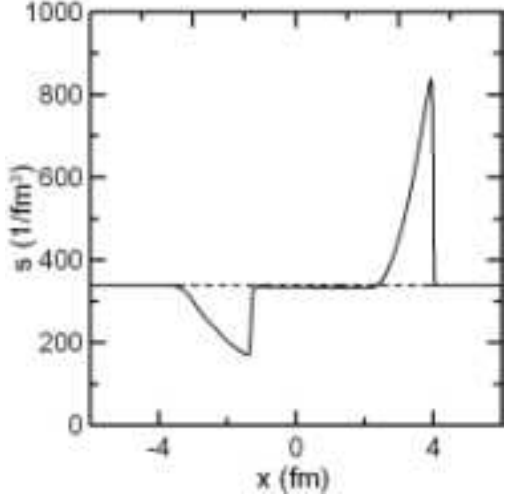}
\caption{The same as Fig. \protect\ref{fig12a}, calculated with the
additional viscosity. }
\label{fig12b}
\end{figure}

Another typical example where shock discontinuities appear is that when a
component of fluid is accelerated by an external force and achieves a
velocity greater than that of the sound in the medium. Such phenomena are
particularly interesting in heavy ion physics. When high energy partonic
jets punch out the thermally equilibrated QGP, they may transfer a large
amount of momentum and energy, dragging the piece of the fluid with high
velocity. Such a scenario is often discussed with the observed angular
correlation of produced hadrons, connecting to the formation of Mach cones%
\cite{Mach, 1st-b}. Of course, the Mach cones do not exist in 1+1
dimensional systems, but it is interesting to study the dynamical formation
of a shock discontinuity in our theory. If we calculate such situation
without the additional viscosity, the formation of shock leads to quick
unphysical oscillations as shown in Fig. \ref{fig12a}. Here, a small part of
the fluid has an finite initial velocity (Lorentz factor $\gamma \approx 2$)
with constant entropy density in the local rest frame. However as is
discussed in the previous section, the inclusion of the additional viscosity
can eliminate these unphysical oscillations as shown in Fig. \ref{fig12b},
confirming the efficiency of our additional viscosity. Differently from the
case of discontinuity initial condition shown in Fig. \ref{fig4}, we found
that the shock front created by a rapidly moving fluid element cannot be
smoothed out even in the presence of normal viscosity. The effect of piling
up the matter at the shock front generates a steep density variation whose
wavelength becomes eventually smaller than the coarse-graining scale $h.$
Therefore, the introduction of the additional viscosity is essential to
calculate the dynamical shock formation processes.

\begin{figure}[tbp]
\includegraphics[scale=1]{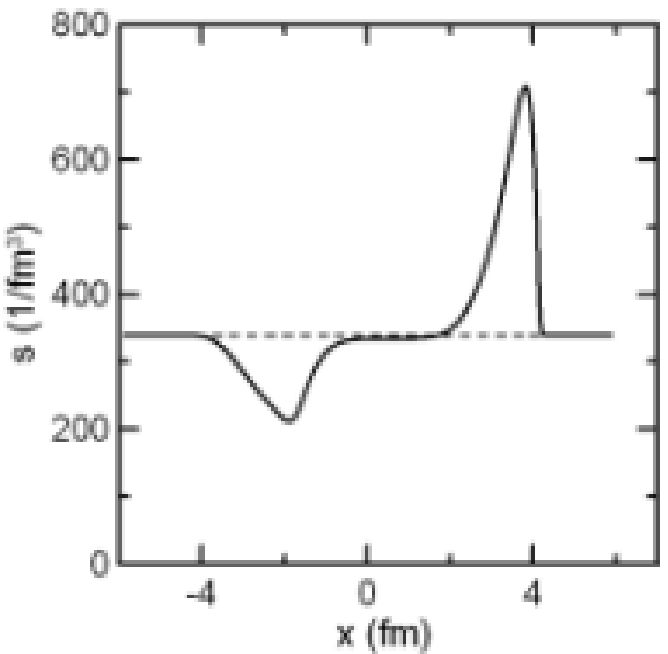}
\caption{The same as Fig. \protect\ref{fig12b} with $a=0.1$.}
\label{fig13}
\end{figure}

\begin{figure}[tbp]
\includegraphics[scale=1]{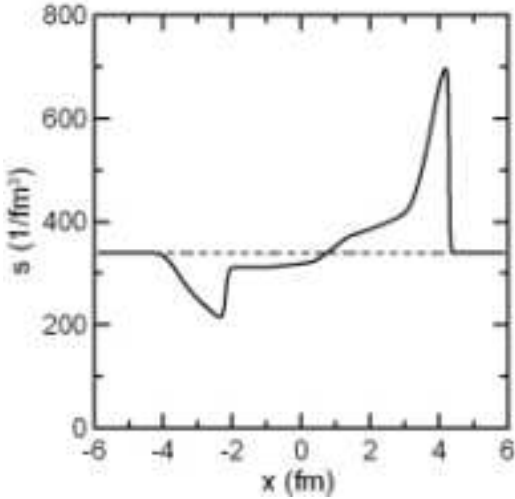}
\caption{The same as Fig. \protect\ref{fig12b} with $a=1$.}
\label{fig14}
\end{figure}

\begin{figure}[tbp]
\includegraphics[scale=1]{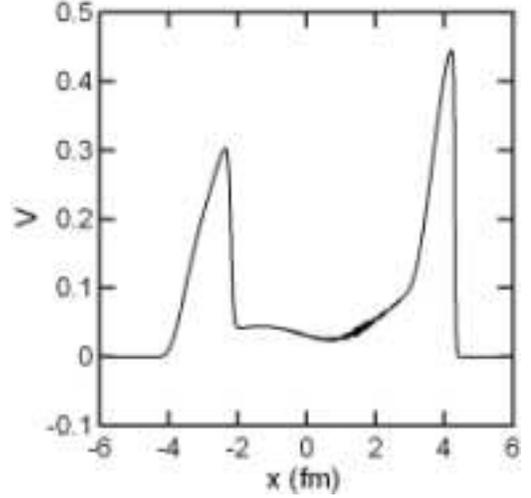}
\caption{The corresponding velocity profile of the viscous fluid of $a=1$.}
\label{fig15}
\end{figure}

In Figs. \ref{fig13} and \ref{fig14}, we show the entropy density profile of
the same situation for different values of viscous coefficient $a$ with
additional viscosity. It is interesting to note that a rarefaction wave is
formed and propagates backwards with a smaller velocity, as is shown in
these figures, although the fluid velocity is positive everywhere, as is
shown in Fig. \ref{fig15}. Note that, another density discontinuity is
formed in the rarefaction wave part which behaves exactly in the same manner
as the case of shock front created by the initial density discontinuity.

\section{Discussion and concluding remarks}

In this paper, we studied the causal dissipative hydrodynamics in 1+1
dimensional systems. To clarify the effects of viscosity and its relation to
the coarse-graining aspect of the theory, we apply the SPH formulation to
represent the macroscopic variables. Here, we consider the SPH scale
parameter $h$ as the scale of coarse-graining.

We have shown that once the viscosity is introduced appropriately, then the
expansion of the fluid into the vacuum should form a steady wave and there
the universal relation $p+\Pi =0$ should be satisfied. We also studied the
various situations where the shock discontinuities emerge. We argue that for
the consistency of the causal dissipative hydrodynamics, we have to
introduce an additional viscosity associated with the coarse-graining scale
of the theory. This is because, the normal viscosity obtained from
microscopic theories, such as kinetic equations or the Green-Kubo-Nakano
formula, is associated with the bulk properties of the matter and is
calculated in the vanishing coarse-graining scale $h\rightarrow 0.$ To be
consistent with the coarse-graining scheme of the theory, there should be
additional dissipation mechanisms that disappears in the limit of vanishing $%
h.$ We proposed a scheme which contemplates such an additional viscosity,
keeping causality and stability of the theory.

Most of the results in this study will be more relevant when the role of
viscosity becomes effective. Such situations will be realized in the coming
LHC experiments where we may expect that a large inhomogeneity in the
velocity profile can be created. The application of the present theory for
the LHC energy is now in progress.

One might think that the problems discussed here might be out of the range
of applicability of hydrodynamics in the usual argument based on the
smallness of the Knudsen number. However, as we already pointed out, it is
well-known that hydrodynamics is still applied even for cases where the
deviation from equilibrium is apparently large. In fact, the microscopic
derivation of hydrodynamics is still an open question. Interesting works are
still under progress to justify the applicability of hydrodynamics based on
the asymptotic theory, the fluctuation theory and so on. In our opinion,
while we do not establish a precise theoretical criteria for its use,
hydrodynamics should be explored independent of such formal limitations,
since it contains very important ingredients to describe phenomenologically
the collective aspects of matter. As a matter of fact, if we had really to
restrict ourselves to the region where the hydrodynamics approach is clearly
available, we would never discuss the phase transition dynamics in this
scheme \cite{koide3}, so that all the ideal fluid model for heavy-ion
collisions would become meaningless. In the same sense, we consider that it
is significant to study the causal dissipative hydrodynamics comprehensively.

In this paper, we discussed the problem of causality and stability of our
theory, and concluded that the theory is stable for the small linear
perturbations. To be precise, we did not prove that the theory is stable for
nonlinear perturbations, although the numerical solutions examined here did
not show any instabilities. However, if the theory is shown to be unstable
in some regime, then we have to construct another stable theory. Analysis on
this line is under investigation and will be reported in another paper. The
effect of the equation of state containing phase transitions also will be
discussed in future.

This work has been supported by CNPq, FAPERJ, CAPES and PRONEX.

\appendix

\section{Telegraph equation}

We discuss the behavior of the causal diffusion equation, which is described
by 
\begin{equation*}
\tau \frac{\partial ^{2}}{\partial t^{2}}n+\frac{\partial }{\partial t}n-D%
\frac{\partial ^{2}}{\partial x^{2}}n=0.
\end{equation*}%
As is well-known, the maximum propagation speed of signals described by the
equation is $v=\sqrt{D/\tau }$. As is discussed in \cite{MF}, the solution
of the causal diffusion equation is given by 
\begin{eqnarray}
n(x,t) &=&\frac{1}{2}e^{-\frac{t}{2\tau }}\left[ n_{0}(x+vt)+n_{0}(x-vt)%
\right]  \notag \\
&& \hspace{-1cm} +e^{-\frac{t}{2\tau }}\int_{x-vt}^{x+vt}dx_{0}\left\{ \sqrt{%
\frac{1}{D\tau }}\frac{1}{4}I_{0}\left[ \frac{1}{2}\sqrt{\frac{1}{D\tau }}%
\sqrt{v^{2}t^{2}-(x-x_{0})^{2}}\right] \right.  \notag \\
&& \hspace{-1cm} \left. +\frac{1}{2}\sqrt{\frac{\tau }{D}}\frac{\partial }{%
\partial t}I_{0}\left[ \frac{1}{2}\sqrt{\frac{1}{D\tau }}\sqrt{%
v^{2}t^{2}-(x-x_{0})^{2}}\right] \right\} n_{0}(x_{0})  \notag \\
&& \hspace{-1cm} +\frac{1}{2}\sqrt{\frac{\tau }{D}}e^{-\frac{t}{2\tau }%
}\int_{x-vt}^{x+vt}dx_{0}I_{0}\left[ \frac{1}{2}\sqrt{\frac{1}{D\tau }}\sqrt{%
v^{2}t^{2}-(x-x_{0})^{2}}\right]  \notag \\
&&\hspace{-1cm} \times \frac{\partial }{\partial t}n_{0}(x_{0}),
\end{eqnarray}%
where $n_{0}(x)$ is the initial distribution and $I_{0}(x)$ is the modified
Bessel function. The first term of the solution indicates that the initial
distribution $n_{0}(x)$ is separated into two fragments, which travel to the
opposite directions with the velocity $v$. Thus, in the shorter time scale
than the relaxation time $\tau $, the memory of the initial distribution
profile survives near the boundary.

\end{document}